\documentclass[12pt,preprint]{aastex}

\usepackage{amsmath}
\usepackage{amssymb}

\begin{document}

\title{Nuclear Partition Functions at Temperatures Exceeding 10$^{10}$ K}


\author{T.\ Rauscher}
\affil{Departement f\"ur Physik und Astronomie, Universit\"at Basel, 4056
Basel, Switzerland}
\email{Thomas.Rauscher@unibas.ch}


\begin{abstract}


Nuclear partition functions were calculated for a grid of temperatures
from $1.2\times 10^{10}$ K to $2.75\times 10^{11}$ K 
($1\leq kT \leq 24$ MeV) within a 
Fermi-gas approach, including all nuclides from the proton-dripline to
the neutron-dripline with proton number $9\leq Z\leq 85$.
The calculation is based on
a nuclear level density description published elsewhere, thus extending
the previous tables of partition functions beyond $10^{10}$ K. Additional
high temperature corrections had to be applied.

\end{abstract}


\section{Introduction}
\label{sec:intro}

The knowledge of the nuclear partition function at high temperatures is
essential in understanding the nuclear equation of state used in the core 
collapse phase of massive stars.
In self-consistent simulations, the contraction of the core is
explicitly followed up to nuclear densities, giving rise to extreme
temperatures and high mean excitation energies of the nuclei.
Ratios of high temperature partition functions are also ingredients
in nucleosynthesis networks in explosive scenarios, such as the
r- and rp-processes. When employed in nuclear statistical
equilibria (NSE), they often have to be known at temperatures
beyond $10^{10}$ K.

Recently, new sets of partition functions have been published along
with astrophysical reaction rates for nuclides
from proton dripline to neutron dripline
and charge number $10\leq Z\leq 85$ \citep{RATH}. The sets include
partition functions up to $T_9=10$ ($10^{10}$ K) based on two different
level densities calculated within a shifted Fermi-gas approach \citep{rtk97}
utilizing two mass formulas. Here, the extension of these partition functions
to temperatures of $T_9=275$ is presented. A straightforward extrapolation
is not valid because of additional effects acting at high temperatures.

These effects have been a matter of discussion already about 20 years ago
\citep{fow78,maz79}. The recently improved descriptions of nuclear level
density and nuclear reaction rate predictions make it worthwhile to
reconsider these arguments and to publish a complete table of partition
functions. In this work, in addition to using the more recent level
densities of \citet{rtk97}, the corrections are treated by closely following
\citet{tub79}.

\section{Procedure}
\label{sec:proc}

The temperature-dependent partition function \( G(T) \) normalized
to the ground state spin of the nucleus
\( J^{0} \) is usually defined as \citep{fow67}
\begin{eqnarray}
\label{eq:pfunc}
(2J^{0}+1)G(T)= &  & \sum _{\mu =0}^{\mu _{m}}(2J^{\mu }+1)
e^{-E^{\mu }/kT} \label{eq:partfunc}\\
 &  & +\int\limits _{E^{\mu _{m}}}^{E^{\mathrm{max}}}\sum _{J^{\mu },
\pi ^{\mu }}(2J^{\mu }+1)
e^{-\epsilon/kT}\rho (\epsilon ,J^{\mu },\pi ^{\mu })d\epsilon
\quad ,\nonumber
\end{eqnarray}
with $\rho$ being the level density and \( \mu  \)\( _{m} \) the label
of the last included experimentally known state. The sum over 
Boltzmann-weighted discrete states
from the ground state to state $\mu_m$ is performed using
experimental levels as listed in \citet{rt01}.
Above the last known state an integration over the nuclear level
density is used instead of a summation, as also outlined in \citet{RATH},
employing the level density description of \citet{rtk97}.

The upper limit \( E^{\mathrm{max}} \) of the integration requires
special consideration. Formally, the integration procedure should encompass
energies up to infinity. However, for all practical purposes an energy
cut-off can be introduced because the Boltzmann-factor $e^{-\epsilon/kT}$
dominates at high energies and suppresses any further contributions to the
integral value. It is well known that, for instance,
the maximum excitation energy
above which there are no more significant contributions to the partition
function is of the order of \( 20-25 \) MeV up to $T_9=10$ \citep{RATH}.

\begin{figure}
\plotone{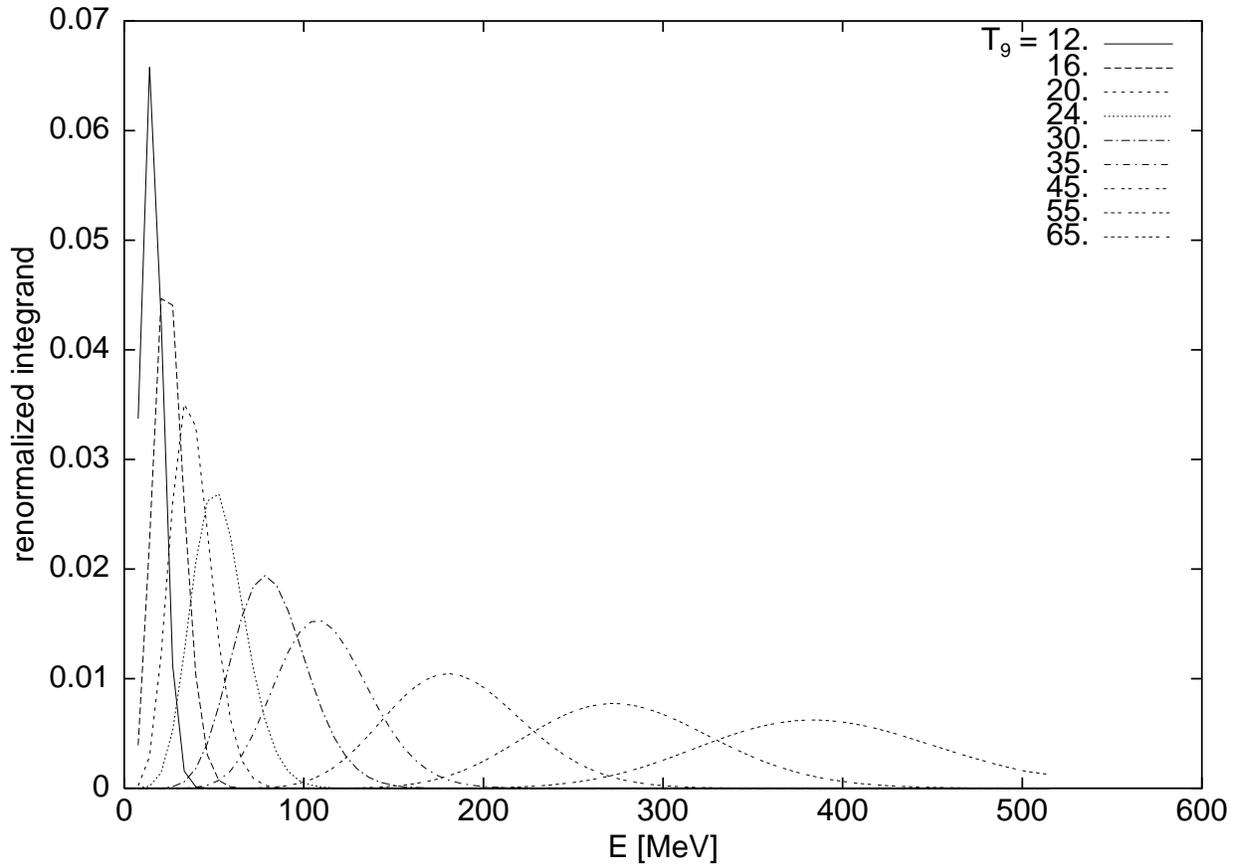}
\figcaption[f1.ps]{\label{fig:peaks}Integrands from 
Eq.\ \protect\ref{eq:pfunc} for different temperatures $T_9$ of $^{109}$Cd. 
The absolute
values are renormalized so that the area under the curves is the same.
It can be seen that for increasing temperature the location of the peak,
arising from folding the Boltzmann factor $e^{-E/kT}$ with the level density
$\rho(E)$, is shifted to increasingly higher excitation energies $E$.
At the same time, the width of the peak is increased, thus allowing
significant contributions to the integral at even higher energies.}
\end{figure}

Due to the temperature dependence of the integrand in Eq.\ \ref{eq:pfunc}
its peak contribution
is shifted to higher energies for higher temperatures $T$, thus also
requiring a larger cut-off \( E^{\mathrm{max}} \). Up to now, there has
been no systematic scrutiny of the behavior of the integrand, which also
weakly depends on the used level density. In Fig.\ \ref{fig:peaks},
the integrands are plotted, also showing the peak energies and the widths
of the peaks
for different energies. The shown energies are in agreement with the mean
excitation energies derived by \citet{tub79}. In the same manner, the
cut-off energy of 25 MeV, often used for calculating partition functions
up to $T_9=10$, can be justified.

For $T_9>12$, we extract a (nearly) quadratic dependence on
temperature of the peak energy $E_{\rm peak}$ and a linear dependence
of the width $\Gamma_\mathrm{FWHM}$ of the integrand:
\begin{eqnarray}
\label{eq:peak}
E_{\rm peak}&=&0.0725 T_9^{2.055} \quad {\rm MeV}, \\
\Gamma_\mathrm{FWHM} &=& 3 T_9-37.0 \quad {\rm MeV}. \nonumber
\end{eqnarray}
The integration cut-off was then set to \( E^{\mathrm{max}} = \max
\left(35,E_{\rm peak} + \Gamma_\mathrm{FWHM} \right) \) MeV.

\section{High temperature corrections}

Due to the exponential increase of the nuclear level density with
excitation energy, extremely
large partition functions already result at temperatures of a few 
MeV (temperatures given as energies and in $T_9$ are related by 
$E=T_9/11.6045$ MeV).
However, it has been realized that a straightforward integration over the
level density might overestimate the partition functions. High excitation
energies of the nucleus permit the emission of nucleons and therefore
an appropriate fraction of the level density associated with such
continuum states should be neglected in the computation of the partition
function.

\citet{fow78} introduced such high temperature corrections by truncating the
integration near the nucleon separation energy and by subtracting
continuum scattering states (which, however, do not act below $T_9=100$).
\citet{maz79} accounted for the suppression of the partition functions
by arbitrarily setting the integral cut-off to 25 MeV. In a semi-classical
calculation, \citet{tub79} showed that \citet{fow78} and \citet{maz79}
largely overestimated the suppression, that a simple truncation of the integral
is incorrect, and that partition functions remain large for temperatures
as high as $T_9=100$. They find that the corrections are much smaller than
given by truncated level density integrals and that the conventional
partition functions (with full integration) are much closer to their values
than values obtained with any of the truncation methods.

The advantage of the description by \citet{tub79}, which is based on the
independent particle model, is the natural inclusion of both bound and
continuum nuclear states. Here, we use a hybrid model by using the
level density and partition function descriptions as outlined in Sec.\
\ref{sec:proc} and applying correction factors derived from the spherical
square well approximation of
\citet{tub79} (Eqs.\ 7 and 9 in that reference)
but using the same nuclear properties (nucleon separation
energies, nuclear radius) as in \citet{RATH}. This way, a continuous
extension of the partition functions of \citet{RATH} is possible. While
the simplicity of the equations is kept, the
limitations of the spherical square well approach are partially lifted
because, e.g., the separation energies are taken from experiment or
from mass formulas employing
more realistic nuclear potentials and accounting for shell and
deformation effects. Furthermore, this approach is only used to
obtain the relative corrections.

The correction factor $C$ is extracted by comparing the uncorrected and
the corrected total nuclear partition function of \citet{tub79} computed
in their spherical square well formalism.
While referring the reader to the paper of \citet{tub79} for a more complete
description of their approach, only the relevant equations are summarized
here. The total nuclear partition function $Z=Z_\mathrm{esw}
=(2J_{i}^{0}+1)G_\mathrm{esw}(T)$
is constructed as the sum of two terms for protons and neutrons, respectively:
\begin{equation}
\ln Z = \ln Z_p + \ln Z_n \quad ,
\end{equation}
with
\begin{equation}
\label{eq:nucleonpart}
\ln Z_x = \ln q_x - \alpha X + \beta E_{0x} - \frac{1}{2} 
\ln (2\pi \mathfrak{N})\quad.
\end{equation}
The letter $x$ stands for neutron (n) and proton (p), respectively, and $X$
is the neutron number $N$ and the proton number $Z$, respectively. The ground
state energy is denoted by $E_{0x}$ and the inverse nuclear temperature
by $\beta=1/kT$ with $\beta=11.6045/T_9$ MeV. The mean-square number
fluctuation $\mathfrak{N}$, the nuclear contribution  $q_x$
(as opposed to the contribution of the exterior nucleon gas) 
of the grand partition function,
and the Lagrange multiplier $\alpha$ can be found with and without
continuum contributions, leading to nucleon partition functions $Z_x$, $Z_x'$
and total partition functions
$Z$, $Z'$ with and without corrections. In the following, primed quantities
are without corrections. Thus, we obtain
\begin{eqnarray}
q_x'=D(T)\left[F_{3/2}\left(\alpha'+\beta S_x+3\beta X/2\rho_F\right)\right]
\quad,\nonumber \\
q_x=D(T)\left[F_{3/2}\left(\alpha+\beta S_x+3\beta X/2\rho_F\right)-
F_{3/2}\left(\alpha\right)\right] \quad,
\end{eqnarray}
and
\begin{eqnarray}
\mathfrak{N}'=\frac{3}{4}D(T)\left[F_{-1/2}\left(\alpha'+\beta S_x+3\beta X/2\rho_F\right)\right]
\quad,\nonumber \\
\mathfrak{N}=\frac{3}{4}D(T)\left[F_{-1/2}\left(
\alpha+\beta S_x+3\beta X/2\rho_F\right)-
F_{-1/2}\left(\alpha\right)\right] \quad.
\end{eqnarray}
Fermi integrals of the order $\eta$ with argument $\theta$
are denoted by $F_\eta\left(\theta\right)$.
The factor $D(T)$ is
\begin{equation}
D(T)=\frac{1}{\sqrt{X}}\left( \frac{2\rho_F kT}{3}\right)^\frac{3}{2} \quad.
\end{equation}
For consistency, the same particle separation energy $S_x$ is used as 
for the reaction rate calculations of \citet{RATH}. It is taken either
from experiment or from a mass formula where no experimental information
is available. The level density at the
zero-temperature Fermi surface is given as 
\begin{equation}
\rho_F=\left( \frac{4}{\sqrt{3\pi}}\right) ^\frac{2}{3}X^\frac{1}{3} \frac{m_x R^2}{\hbar^2}
\quad,
\end{equation}
using the nuclear radius $R$ and the nucleon mass $m_x$. With that definition
the ground state energy becomes
\begin{equation}
E_{0x}=-\frac{3}{5}\frac{X^2}{\rho_F}-XS_x \quad.
\end{equation}

Before evaluating the above equations, the appropriate 
(temperature dependent) Lagrange multiplicators
with and without corrections have to be determined. This is done by 
requiring states in the grand canonical ensemble to have, on the average,
the correct number of nucleons, $X$, and therefore by finding
the root of the following equations with respect to $\alpha$ and $\alpha'$:
\begin{eqnarray}
&\frac{3}{2}D(T)\left[F_{1/2}\left(\alpha'+\beta S_x+3\beta X/2\rho_F\right)
\right]-X&=0
\quad,\nonumber \\
&\frac{3}{2}D(T)\left[F_{1/2}\left(\alpha+\beta S_x+3\beta X/2\rho_F\right)-
F_{1/2}\left(\alpha\right)\right] -X&=0\quad.
\end{eqnarray}
The proper $\alpha$ or $\alpha'$ found above
has to be inserted also in Eq.\ \ref{eq:nucleonpart}, of course.

Finally, the relevant partition function $\overline{G}(T)$ is then obtained by
multiplying the previous function (from Sec.\ \ref{sec:proc})
with the correction $C$:
\begin{equation}
\overline{G}(T)=C(T)G(T)=\exp \left( \ln Z(T)- \ln Z'(T) \right) G(T)\quad.
\end{equation}
Thus, the correction factor $C$ found with the approach above is applied
to the partition function derived in the full computation described in
Sec.\ \ref{sec:proc}.
The corrections start to act
at temperatures $T_9\simeq 50-60$ for light and intermediate nuclei and as
low as $T_9\simeq 14$ for heavy nuclei. Corrections are negligible for
$T_9\leq 10$, implying that the partition functions from \citet{RATH} can
be used without further modifications. The magnitude of the corrections
ranges from a few percent at the lower end of the temperature range to
a suppression factor of $10^{-5}$ for the heaviest nuclides at $T_9=100$
and $10^{-40}$ at $T_9=275$, respectively.
The correction factors for a few selected cases are shown in Table
\ref{tab:corr}.

\section{Discussion and Conclusion}

The corrected renormalized
partition functions $\overline{G}$ calculated with level densities utilizing
input from the Finite Range Droplet Model (FRDM)
\citep{mol95} (see also Rauscher \&
Thielemann 2000) are given in Table \ref{tab:pf_frdm}.
Results making use of the Extended Thomas-Fermi mass formula with shell
quenching effects (ETFSI-Q) \citep{abuxx} (see also Rauscher \& Thielemann 2000)
far from stability are given in Table \ref{tab:pf_etfsiq}.
The properties of the mass formulas can enter via the particle
separation energies which are calculated from predicted mass differences
in case no experimental masses are known. Furthermore, they always enter in the
microscopic correction term used in the level density treatment of
\citet{rtk97}. The method to calculate the high temperature corrections
is only applicable for bound nucleons, therefore only those nuclides are
given for which both the neutron and proton separation energies are
positive.
The printed version of this paper contains only example tables, showing
which kind of information is available. Partition functions for the full 
range of nuclides from
proton dripline to neutron dripline for $10\leq Z\leq 83$ (FRDM) and
$26\leq Z\leq 85$ (ETFSI-Q)
are available as machine readable tables in electronic form.
The formatting is the same as used
in \citet{RATH}, except for the different temperature range. Thus, the
partition functions presented here provide a smooth and analytical extension
of the previous tabulation, extending the range of temperatures to
$0.1\leq T_9\leq 275.$

\begin{figure}
\plotone{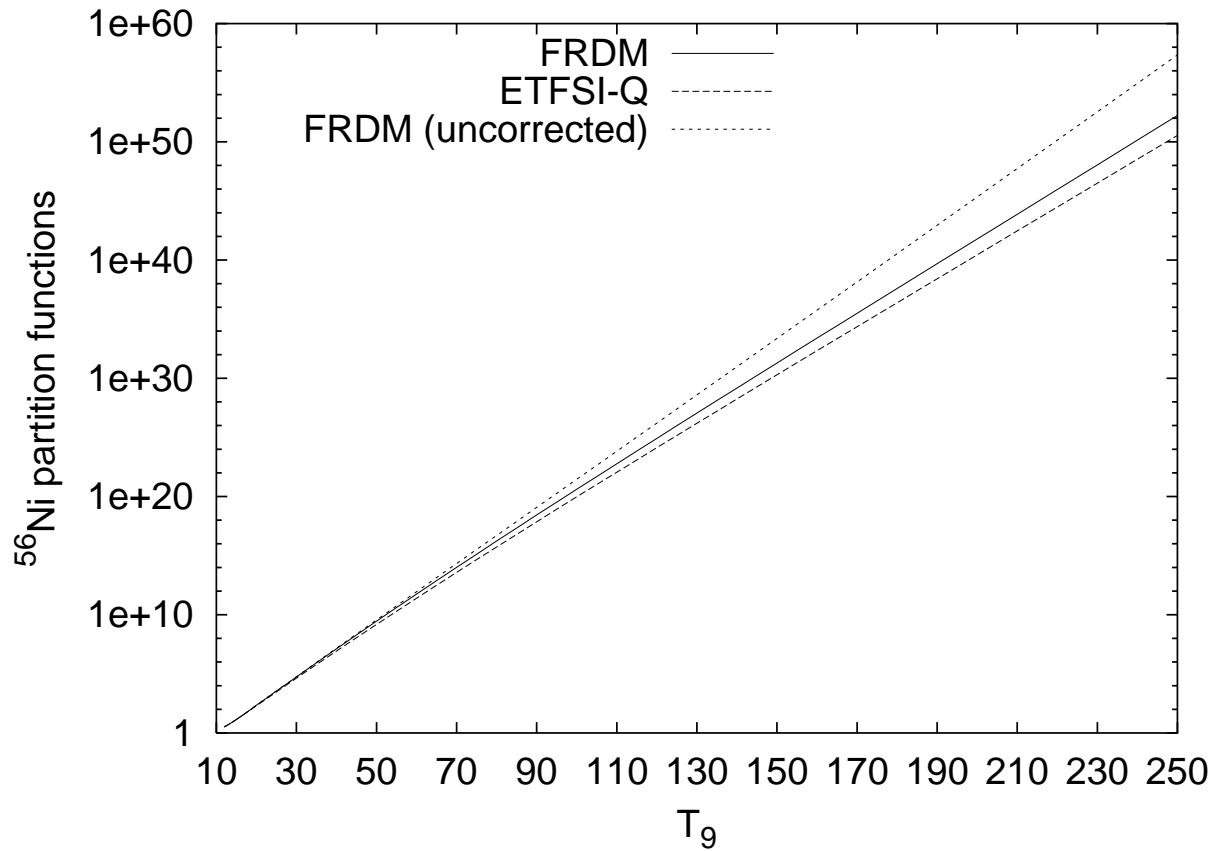}
\figcaption[f2.ps]{\label{fig:ni56}Partition function of $^{56}$Ni 
calculated
with level densities including inputs from FRDM (full line) and ETFSI-Q (dashed
line). Both calculations include high-temperature corrections which, however,
become significant only at $kT>5$ MeV for this nucleus. Differences
between FRDM and ETFSI-Q partition functions are more pronounced for
neutron-rich nuclides. Also shown is a partition function without the
continuum corrections (dotted line).}
\end{figure}

The new values for $^{56}$Ni can directly be compared to the ones from
\citet{tub79}. Fig.\ \ref{fig:ni56} shows the partition function of this
nucleus. By comparing to Fig.\ 1 in \citet{tub79} it can be seen that the
new value is higher by 45--50\% around $kT=10$ MeV than their corrected 
value B. 
This is mainly due to the different level density description (different
effective level density parameter $a$) used since a similar
treatment of the high-temperature corrections is implemented in both
calculations.

It has to be noted that the partition functions presented here are valid
for low-density conditions. In high-density regimes, modifications of
nuclear properties (e.g., separation energies) might have to be
additionally applied. This is beyond the scope of the current
investigation.

The nuclear model for the corrections (and the one for the level
density) assumes a Fermi-gas of independent
nucleons interacting only through a common, spin-independent mean field.
At nuclear temperatures beyond about 30 MeV (i.e.\ $T_9 \geq 350$), the
momentum dependence of the mean field, the excitation of mesonic degrees
of freedom, and the breakdown of the independent particle approximation
become important. This is not relevant for the
temperature range explored here but will necessitate an altogether different
approach when expanding the temperature range beyond about 25--30 MeV.
It is expected that the exponential rise of the partition functions with
temperature will finally be effectively suppressed beyond those energies.

\acknowledgments

This work is supported by
the Swiss National Science
Foundation (2000-061031.02). T.R. acknowledges a PROFIL
professorship from the Swiss National Science foundation (grant
2024-067428.01).





\clearpage



\clearpage

\clearpage

\begin{deluxetable}{rrrrrrrr}
\tabletypesize{\footnotesize}
\tablecaption{\label{tab:corr}Correction factors
$C(T_9)$ for selected cases. 
Numbers in square brackets denote powers of ten.
}
\tablehead{
\colhead{Nuclide} &&&&&&&\\
\colhead{$C(12)$}&
\colhead{$C(14)$}&
\colhead{$C(16)$}&
\colhead{$C(18)$}&
\colhead{$C(20)$}&
\colhead{$C(22)$}&
\colhead{$C(24)$}&
\colhead{$C(26)$} \\
\colhead{$C(28)$}&
\colhead{$C(30)$}&
\colhead{$C(35)$}&
\colhead{$C(40)$}&
\colhead{$C(45)$}&
\colhead{$C(50)$}&
\colhead{$C(55)$}&
\colhead{$C(60)$} \\
\colhead{$C(65)$}&
\colhead{$C(70)$}&
\colhead{$C(75)$}&
\colhead{$C(80)$}&
\colhead{$C(85)$}&
\colhead{$C(90)$}&
\colhead{$C(95)$}&
\colhead{$C(100)$} \\
\colhead{$C(110)$}&
\colhead{$C(130)$}&
\colhead{$C(150)$}&
\colhead{$C(170)$}&
\colhead{$C(190)$}&
\colhead{$C(210)$}&
\colhead{$C(230)$}&
\colhead{$C(250)$}
}
\startdata
$^{16}$O&&&&&&&\\
1.00[$+00$]&1.00[$+00$]&1.00[$+00$]&1.00[$+00$]&1.00[$+00$]&1.00[$+00$]&1.00[$+00$]&1.00[$+00$]\\
9.99[$-01$]&9.99[$-01$]&9.97[$-01$]&9.93[$-01$]&9.87[$-01$]&9.78[$-01$]&9.65[$-01$]&9.49[$-01$]\\
9.29[$-01$]&9.05[$-01$]&8.79[$-01$]&8.49[$-01$]&8.17[$-01$]&7.83[$-01$]&7.48[$-01$]&7.11[$-01$]\\
6.36[$-01$]&4.90[$-01$]&3.63[$-01$]&2.61[$-01$]&1.83[$-01$]&1.27[$-01$]&8.66[$-02$]&5.89[$-02$]\\
\hline
$^{56}$Fe&&&&&&&\\
1.00[$+00$]&1.00[$+00$]&1.00[$+00$]&1.00[$+00$]&9.99[$-01$]&9.98[$-01$]&9.96[$-01$]&9.93[$-01$]\\
9.88[$-01$]&9.82[$-01$]&9.60[$-01$]&9.24[$-01$]&8.75[$-01$]&8.14[$-01$]&7.42[$-01$]&6.64[$-01$]\\
5.83[$-01$]&5.03[$-01$]&4.26[$-01$]&3.55[$-01$]&2.90[$-01$]&2.34[$-01$]&1.86[$-01$]&1.46[$-01$]\\
8.69[$-02$]&2.71[$-02$]&7.50[$-03$]&1.90[$-03$]&4.56[$-04$]&1.06[$-04$]&2.40[$-05$]&5.43[$-05$]\\
\hline
$^{56}$Ni&&&&&&&\\
1.00[$+00$]&1.00[$+00$]&9.99[$-01$]&9.98[$-01$]&9.96[$-01$]&9.92[$-01$]&9.88[$-01$]&9.82[$-01$]\\
9.74[$-01$]&9.65[$-01$]&9.33[$-01$]&8.90[$-01$]&8.35[$-01$]&7.71[$-01$]&7.00[$-01$]&6.24[$-01$]\\
5.48[$-01$]&4.73[$-01$]&4.02[$-01$]&3.36[$-01$]&2.78[$-01$]&2.26[$-01$]&1.81[$-01$]&1.44[$-01$]\\
8.72[$-02$]&2.86[$-02$]&8.33[$-03$]&2.23[$-03$]&5.61[$-04$]&1.36[$-04$]&3.24[$-05$]&7.63[$-06$]\\
\hline
$^{176}$Hf&&&&&&&\\
9.99[$-01$]&9.97[$-01$]&9.93[$-01$]&9.85[$-01$]&9.72[$-01$]&9.52[$-01$]&9.26[$-01$]&8.92[$-01$]\\
8.51[$-01$]&8.04[$-01$]&6.61[$-01$]&5.05[$-01$]&3.57[$-01$]&2.33[$-01$]&1.42[$-01$]&8.07[$-02$]\\
4.29[$-02$]&2.15[$-02$]&1.01[$-02$]&4.54[$-03$]&1.93[$-03$]&7.88[$-04$]&3.08[$-04$]&1.16[$-04$]\\
1.49[$-05$]&1.77[$-07$]&1.55[$-09$]&1.12[$-11$]&7.28[$-14$]&4.51[$-16$]&2.79[$-18$]&1.77[$-20$]\\
\hline
$^{208}$Pb&&&&&&&\\
9.99[$-01$]&9.97[$-01$]&9.93[$-01$]&9.84[$-01$]&9.69[$-01$]&9.48[$-01$]&9.18[$-01$]&8.80[$-01$]\\
8.34[$-01$]&7.80[$-01$]&6.23[$-01$]&4.55[$-01$]&3.03[$-01$]&1.85[$-01$]&1.03[$-01$]&5.32[$-02$]\\
2.53[$-02$]&1.12[$-02$]&4.64[$-03$]&1.80[$-03$]&6.62[$-04$]&2.30[$-04$]&7.61[$-05$]&2.41[$-05$]\\
2.14[$-06$]&1.15[$-08$]&4.30[$-11$]&1.28[$-13$]&3.36[$-16$]&8.33[$-19$]&2.06[$-21$]&5.24[$-24$]
\enddata
\tablecomments{The values given
here were calculated with separation energies based on experiment or
FRDM input (see text).}

\end{deluxetable}

\clearpage

\begin{deluxetable}{rrrrrrrr}
\tablecaption{\label{tab:pf_frdm}Renormalized partition functions
$\overline{G}(T_9)$ including high temperature corrections. The values given
here were calculated with level densities based on FRDM input (see text).
Each nuclide is characterized by its charge
and mass numbers $Z$, $A$, and its ground-state spin $J^0$.
Numbers in square brackets denote powers of ten.}
\tablehead{
\colhead{Nuclide} &&&&&&&\\
\colhead{$Z$}&\colhead{$A$}&\colhead{$J^0$}&&&&&\\
\colhead{$\overline{G}(12)$}&
\colhead{$\overline{G}(14)$}&
\colhead{$\overline{G}(16)$}&
\colhead{$\overline{G}(18)$}&
\colhead{$\overline{G}(20)$}&
\colhead{$\overline{G}(22)$}&
\colhead{$\overline{G}(24)$}&
\colhead{$\overline{G}(26)$} \\
\colhead{$\overline{G}(28)$}&
\colhead{$\overline{G}(30)$}&
\colhead{$\overline{G}(35)$}&
\colhead{$\overline{G}(40)$}&
\colhead{$\overline{G}(45)$}&
\colhead{$\overline{G}(50)$}&
\colhead{$\overline{G}(55)$}&
\colhead{$\overline{G}(60)$} \\
\colhead{$\overline{G}(65)$}&
\colhead{$\overline{G}(70)$}&
\colhead{$\overline{G}(75)$}&
\colhead{$\overline{G}(80)$}&
\colhead{$\overline{G}(85)$}&
\colhead{$\overline{G}(90)$}&
\colhead{$\overline{G}(95)$}&
\colhead{$\overline{G}(100)$} \\
\colhead{$\overline{G}(105)$}&
\colhead{$\overline{G}(110)$}&
\colhead{$\overline{G}(115)$}&
\colhead{$\overline{G}(120)$}&
\colhead{$\overline{G}(125)$}&
\colhead{$\overline{G}(130)$}&
\colhead{$\overline{G}(135)$}&
\colhead{$\overline{G}(140)$} \\
\colhead{$\overline{G}(145)$}&
\colhead{$\overline{G}(150)$}&
\colhead{$\overline{G}(155)$}&
\colhead{$\overline{G}(160)$}&
\colhead{$\overline{G}(165)$}&
\colhead{$\overline{G}(170)$}&
\colhead{$\overline{G}(175)$}&
\colhead{$\overline{G}(180)$} \\
\colhead{$\overline{G}(190)$}&
\colhead{$\overline{G}(200)$}&
\colhead{$\overline{G}(210)$}&
\colhead{$\overline{G}(220)$}&
\colhead{$\overline{G}(230)$}&
\colhead{$\overline{G}(240)$}&
\colhead{$\overline{G}(250)$}&
\colhead{$\overline{G}(275)$}
}
\startdata
$^{56}$Ni&&&&&&&\\
28& 56 & 0.0 &&&&&\\
3.23[$+00$]&8.19[$+00$]&2.37[$+01$]&7.17[$+01$]&2.19[$+02$]&6.64[$+02$]&2.01[$+03$]&6.08[$+03$]\\
1.83[$+04$]&5.52[$+04$]&8.60[$+05$]&1.31[$+07$]&1.96[$+08$]&2.86[$+09$]&4.06[$+10$]&5.63[$+11$]\\
7.64[$+12$]&1.02[$+14$]&1.33[$+15$]&1.71[$+16$]&2.17[$+17$]&2.71[$+18$]&3.35[$+19$]&4.10[$+20$]\\
4.96[$+21$]&5.94[$+22$]&7.05[$+23$]&8.30[$+24$]&9.71[$+25$]&1.13[$+27$]&1.30[$+28$]&1.50[$+29$]\\
1.71[$+30$]&1.94[$+31$]&2.20[$+32$]&2.49[$+33$]&2.80[$+34$]&3.14[$+35$]&3.51[$+36$]&3.93[$+37$]\\
4.87[$+39$]&6.02[$+41$]&7.41[$+43$]&9.11[$+45$]&1.12[$+48$]&1.38[$+50$]&1.70[$+52$]&2.90[$+57$]\\
\enddata

\tablecomments{The complete version of this table can be found in the 
electronic edition of Astrophysical Journal Supplement.  
The printed edition contains only
a sample of what kind of information is given. The full tables can also
be downloaded from \url{http://ftp.nucastro.org/astro/fits/partfuncs/}.}

\end{deluxetable}

\clearpage

\begin{deluxetable}{rrrrrrrr}
\tablecaption{\label{tab:pf_etfsiq}Renormalized partition functions
$\overline{G}(T_9)$ including high temperature corrections. The values given
here were calculated with level densities based on ETFSI-Q input (see text).
Each nuclide is characterized by its charge
and mass numbers $Z$, $A$, and its ground-state spin $J^0$.
Numbers in square brackets denote powers of ten.}
\tablehead{
\colhead{Nuclide} &&&&&&&\\
\colhead{$Z$}&\colhead{$A$}&\colhead{$J^0$}&&&&&\\
\colhead{$\overline{G}(12)$}&
\colhead{$\overline{G}(14)$}&
\colhead{$\overline{G}(16)$}&
\colhead{$\overline{G}(18)$}&
\colhead{$\overline{G}(20)$}&
\colhead{$\overline{G}(22)$}&
\colhead{$\overline{G}(24)$}&
\colhead{$\overline{G}(26)$} \\
\colhead{$\overline{G}(28)$}&
\colhead{$\overline{G}(30)$}&
\colhead{$\overline{G}(35)$}&
\colhead{$\overline{G}(40)$}&
\colhead{$\overline{G}(45)$}&
\colhead{$\overline{G}(50)$}&
\colhead{$\overline{G}(55)$}&
\colhead{$\overline{G}(60)$} \\
\colhead{$\overline{G}(65)$}&
\colhead{$\overline{G}(70)$}&
\colhead{$\overline{G}(75)$}&
\colhead{$\overline{G}(80)$}&
\colhead{$\overline{G}(85)$}&
\colhead{$\overline{G}(90)$}&
\colhead{$\overline{G}(95)$}&
\colhead{$\overline{G}(100)$} \\
\colhead{$\overline{G}(105)$}&
\colhead{$\overline{G}(110)$}&
\colhead{$\overline{G}(115)$}&
\colhead{$\overline{G}(120)$}&
\colhead{$\overline{G}(125)$}&
\colhead{$\overline{G}(130)$}&
\colhead{$\overline{G}(135)$}&
\colhead{$\overline{G}(140)$} \\
\colhead{$\overline{G}(145)$}&
\colhead{$\overline{G}(150)$}&
\colhead{$\overline{G}(155)$}&
\colhead{$\overline{G}(160)$}&
\colhead{$\overline{G}(165)$}&
\colhead{$\overline{G}(170)$}&
\colhead{$\overline{G}(175)$}&
\colhead{$\overline{G}(180)$} \\
\colhead{$\overline{G}(190)$}&
\colhead{$\overline{G}(200)$}&
\colhead{$\overline{G}(210)$}&
\colhead{$\overline{G}(220)$}&
\colhead{$\overline{G}(230)$}&
\colhead{$\overline{G}(240)$}&
\colhead{$\overline{G}(250)$}&
\colhead{$\overline{G}(275)$}
}
\startdata
$^{56}$Ni&&&&&&&\\
  28& 56&  0.0&&&&&\\
3.20[$+00$]&8.03[$+00$]&2.28[$+01$]&6.76[$+01$]&2.01[$+02$]&5.94[$+02$]&1.74[$+03$]&5.10[$+03$]\\
1.48[$+04$]&4.30[$+04$]&6.08[$+05$]&8.43[$+06$]&1.14[$+08$]&1.52[$+09$]&1.97[$+10$]&2.50[$+11$]\\
3.11[$+12$]&3.80[$+13$]&4.56[$+14$]&5.40[$+15$]&6.30[$+16$]&7.25[$+17$]&8.25[$+18$]&9.29[$+19$]\\
1.04[$+21$]&1.14[$+22$]&1.25[$+23$]&1.36[$+24$]&1.47[$+25$]&1.57[$+26$]&1.67[$+27$]&1.77[$+28$]\\
1.87[$+29$]&1.96[$+30$]&2.06[$+31$]&2.14[$+32$]&2.23[$+33$]&2.31[$+34$]&2.38[$+35$]&2.46[$+36$]\\
2.60[$+38$]&2.74[$+40$]&2.88[$+42$]&3.02[$+44$]&3.17[$+46$]&3.32[$+48$]&3.50[$+50$]&4.02[$+55$]
\enddata

\tablecomments{The complete version of this table can be found in the 
electronic edition of Astrophysical Journal Supplement.  
The printed edition contains only
a sample of what kind of information is given. The full tables can also
be downloaded from \url{http://ftp.nucastro.org/astro/fits/partfuncs/}.}

\end{deluxetable}

\end{document}